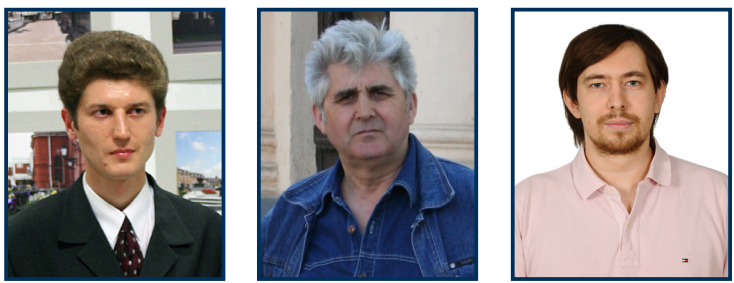

# THE POTENTIAL OF MOBILE EXHIBITION AS A FORM OF IMPLEMENTATION FOR SOCIAL TRANSFORMATION AND EDUCATIONAL EXPO-DESIGN IN ADDRESSING PUBLIC SOCIAL PROBLEMS

*Rifkat I. Nabiyev*
*Department of Fine Art and Costume Art,*
*Faculty of Design and National Culture,*
*Ufa State University of Economics and Service, 450068 Ufa, Russia*
*E-mail: dizain55@yandex.ru*

*Ilshat H. Nabiyev*
*Ufa College of Arts, Ufa, Russia*
*E-mail: dizain55@yandex.ru*

*Rushan Ziatdinov*
*Department of Computer and Instructional Technologies,*
*Fatih University, 34500 Buyukcekmece,*
*Istanbul, Turkey E-mail: rushanziatdinov@yandex.ru*
*URL: http://www.ziatdinov-lab.com/*

# ПЕРЕДВИЖНАЯ ВЫСТАВКА КАК ФОРМА РЕАЛИЗАЦИИ ОБЩЕСТВЕННО-ПРЕОБРАЗУЮЩЕГО И ВОСПИТАТЕЛЬНОГО ПОТЕНЦИАЛА ЭКСПОДИЗАЙНА В КОНТЕКСТЕ РЕШЕНИЯ СОЦИАЛЬНЫХ ПРОБЛЕМ ОБЩЕСТВА

*Р. И. Набиев*
*Уфимский государственный университет экономики*
*и сервиса, г. Уфа, Россия. E-mail: dizain55@yandex.ru*
*URL: http://nabiyev.mathdesign.ru/*

*И. Х. Набиев*
*Уфимский колледж искусств, г. Уфа, Россия.*
*E-mail: dizain55@yandex.ru*

*Рушан Зиатдинов*
*Университет Фатих, г. Стамбул, Турция*
*E-mail: rushanziatdinov@yandex.ru*
*URL: http://www.ziatdinov-lab.com/*

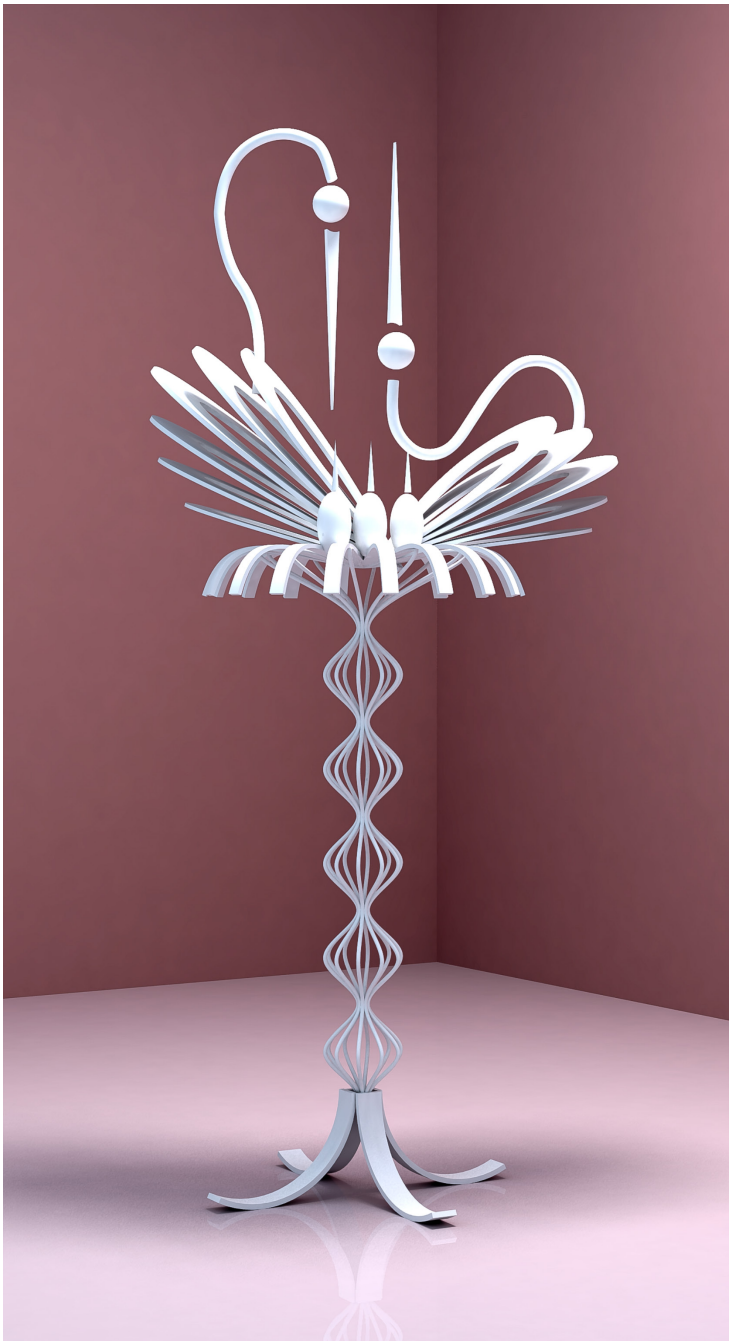
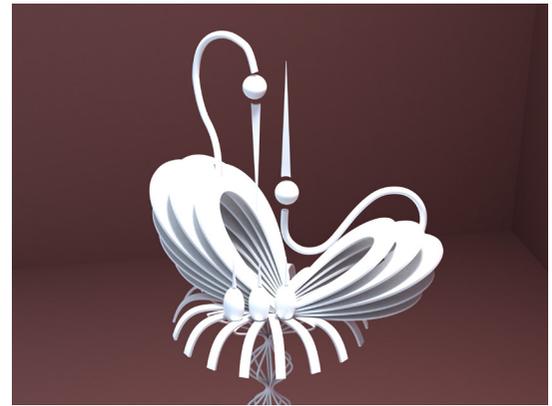
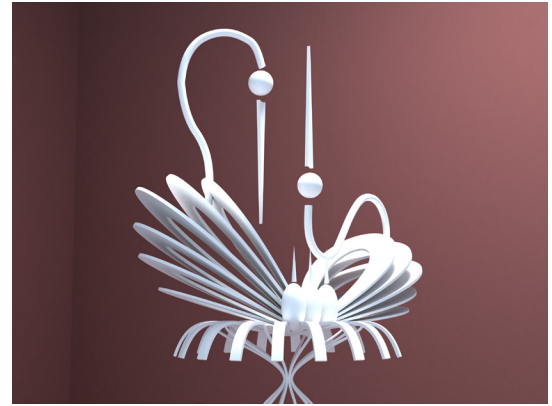
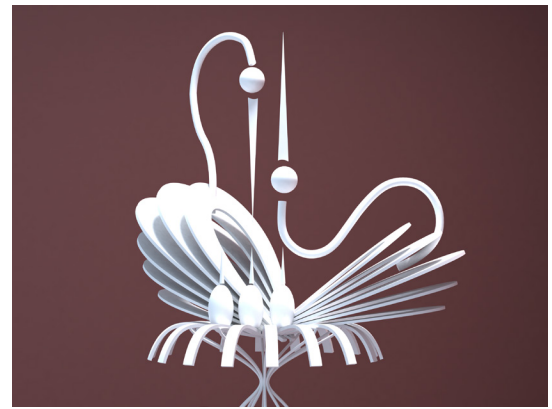

*Рис. 4.*
*Арт-объект.*
*Интерпретация красоты фауны в образе журавлиной семьи.*

***Abstract.*** The article analyzes the humanistic basis of expo-design in conjunction with the ideological problems of modern design and the causes of their occurrence. It reveals the potential of mobile exhibition for social transformation in terms of its structural and ideological aesthetic content. It substantiates the effectiveness of expo-design expression's artistic and aesthetic means in viewer personality moral education. It investigates the features of social problems design presentation in the unity of the formative and design material components features of mobile exhibition.

***Аннотация.*** В статье анализируются гуманистические основы эксподизайна в соотнесенности с идеологическими проблемами современного дизайна и причинами их возникновения. Раскрывается общественно-преобрузующий потенциал передвижной выставки в его структурном и идейно-эстетическом содержании. Обосновывается эффективность художественно-эстетических средств выразительности эксподизайна в нравственном воспитании личности зрителя. Исследуются особенности проектной подачи социальных проблем в единстве формообразующих и конструктивных особенностей материальных компонентов передвижной выставки.

*Ключевые слова:* передвижная выставка; эксподизайн; проектная культура; мировоззрение; проектная идеология; арт-объект; стенд; формообразование; общественно-преобразующая функция.

*Введение*

Существующее состояние духовных и материальных общественных проблем демонстрирует изменения в культурных основах формирования сознания. Все большее распространение получает тенденция привязывать моральные постулаты к воображаемым явлениям, сводить их к формальному нравственному довеску и бутафории повседневной жизни. Примечательно, что в искусстве современные формы отражения существенных для культуры аспектов жизнедеятельности человека как субъекта разнопланового общественного уклада, приняли форму эстетизации мещанского мировоззрения. Культ мнимого «аристократизма» воздвигнут в абсолют, а форма его выражения облична во внешнюю красивость, подтекст которой недвусмысленно выражает направленность на воспитание у зрителя пассивного любования «прелестями безмятежности и неги». Стремление коммерциализировать форму выражения, преподнести ее под выгодным для продажи ракурсом свело на нет всю воспитательную основу искусства. Образ, подчиненный коммерческой выгоде, лишился способности пробуждать стимулы к познанию мира в его диалектическом единстве и противоречии. Беспрепятственно утвердившись в сознании потребителей подобных произведений искусства, такой образ стал мерой нравственной оценки самой действительности.

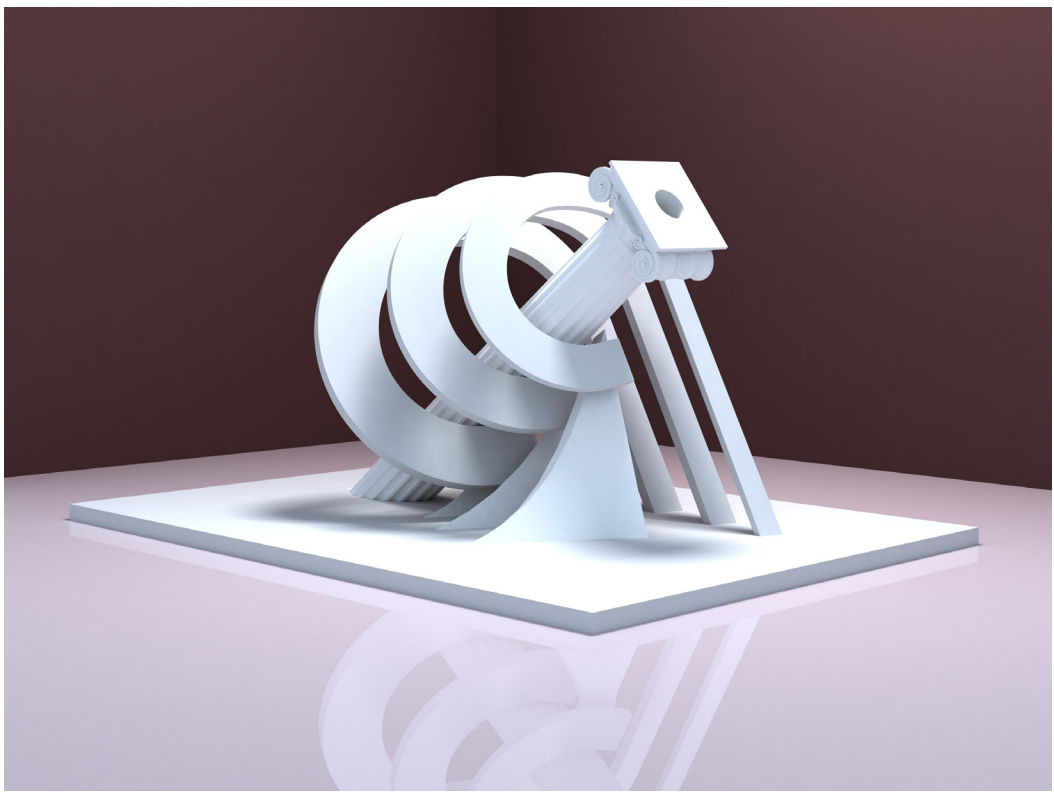

*Рис. 5.*
*Арт-объект передвижной выставки «Архитектура города».*

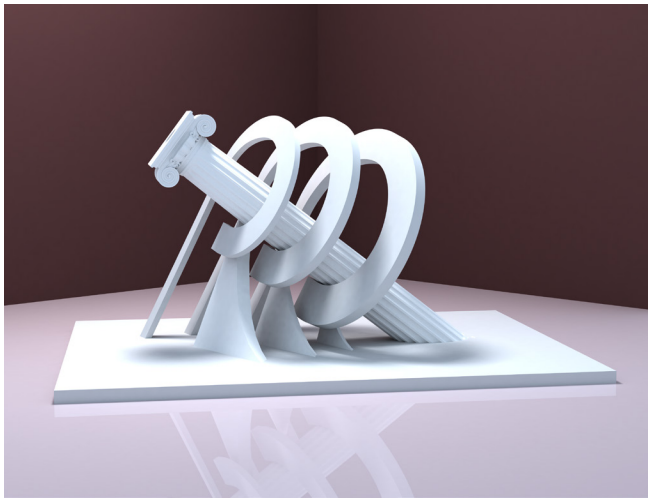 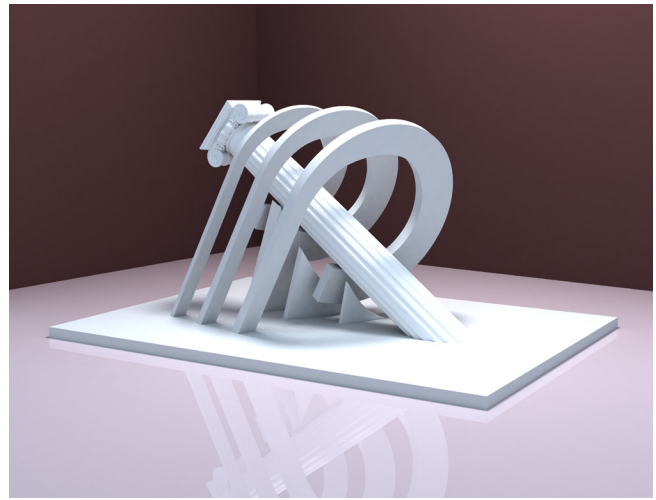

*Рис. 6, 7. Арт-объект передвижной выставки «Архитектура города».*

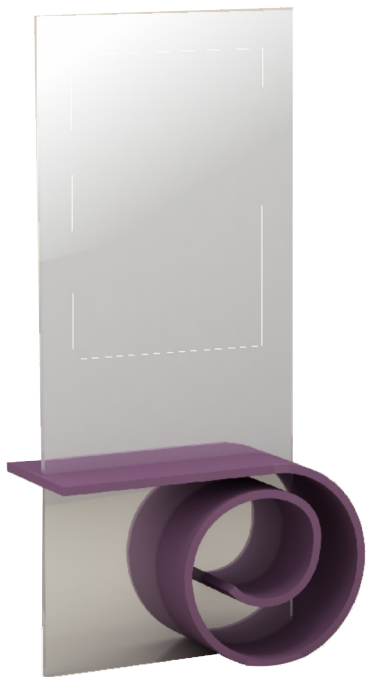

*Рис. 8.*
*Стенд для передвижной вставки «Архитектура города». Изготавливается из пластика и органического стекла. Позволяет демонстрировать плоский наглядный материал с двух сторон. Формообразующие и пластические характеристики сформированы в соответствии со стилевыми особенностями арт-объекта. Выполнен Е. Спиридоновой под руководством Р. И. Набиева.*
*Все проекты арт-объектов и стендов, представленных в данной статье, (кроме обозначенных) являются разработками Р. И. Набиева, И. Х. Набиева, Р. А. Зиатдинова.*

Дизайн как ядро проектной культуры тоже попал под влияние подобного деструктивного процесса. В рамках его видов произошли наглядные изменения. Они, прежде всего, отражают процесс изменения в идеологии профессии, культуре проектной практики и позиции самих дизайнеров в аспекте их отношения к содержанию заказов и социальной ответственности. Многие дизайнеры, в силу известных причин, вынуждены часто занимать соглашательскую позицию относительно художественных вопросов их труда и делать лакейский реверанс перед заказчиками с их «советами». Поэтому, не случайно появилось снисходительное отношение потребителя и заказчика к деятельности дизайнера как к услуге, обслуживающей их капризы.

Известно, что концептуальная сторона профессиональной деятельности художника-конструктора покоится на полноценной школе, которая формирует определенный стиль мышления. В идеале, проектная школа, являясь метасистемой, системой систем, формует конкретные мировоззренческие установки, воспитывает профессиональный вкус для реализации творческих

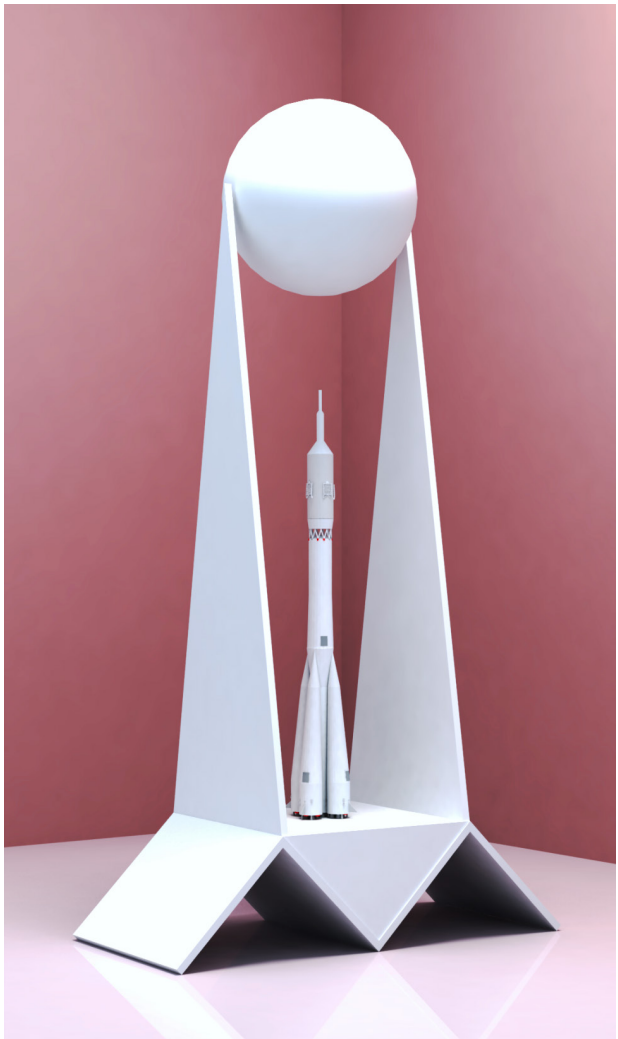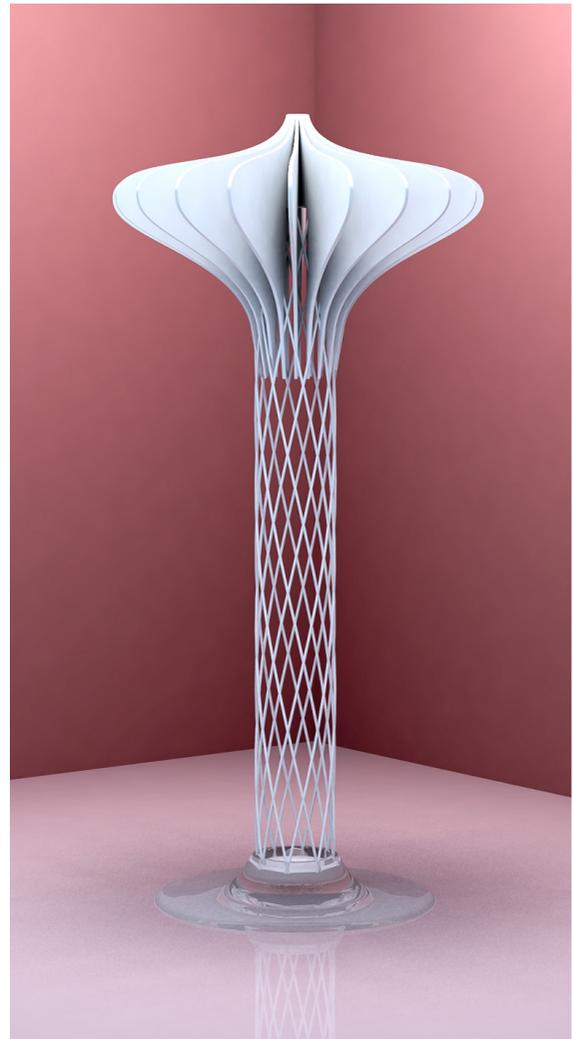

Рис. 9. Арт-объект «Покорение космоса».

Рис. 10. Арт-объект.

потенций в соотнесенности с ценностями профессии. Однако необходимость включения дизайнера в контекст стоящих перед ним проблем на уровне их микроструктуры, а не только формализации идеи в материальных носителях, что удобно и знакомо любому дизайнеру, осознаваема не каждым специалистом.

Причины данной проблемы следует искать в системе подготовки дизайнеров. Ведь не секрет, что в современном дизайнерском образовании очевиден крен в сторону моделирования проектного мышления студента, способного себя проявлять только в рамках мотивации, пробуждаемой коммерческой выгодой. Подобная идеология исключает возникновение интересов, рожденных желанием внести более существенные коррективы в материальную структуру общества мировоззренческого плана. Более того, она порождает объективно-ошибочное представление у субъектов образовательного процесса о сути миссии дизайна в жизни людей. Следование культу «эстетики безобразного» в студенческой среде и преклонение перед кумирами, позиционирующими проектную культуру в данном идейном ключе – лишь некоторые из многих фактов следствия выхолащивания гуманистической сути дизайна из инструментов методического воздействия на обучаемую личность. В результате, в лучшем случае, из выпускника получается ремесленник, пытающийся хоть как-то приложить полученные навыки, в худшем - дезориентированная в идейно-эстетических вопросах профессии личность.

### Содержание передвижной выставки как направления эксподизайна

В проектном пространстве функционирования дизайна обозначились конкретные его виды, ко-

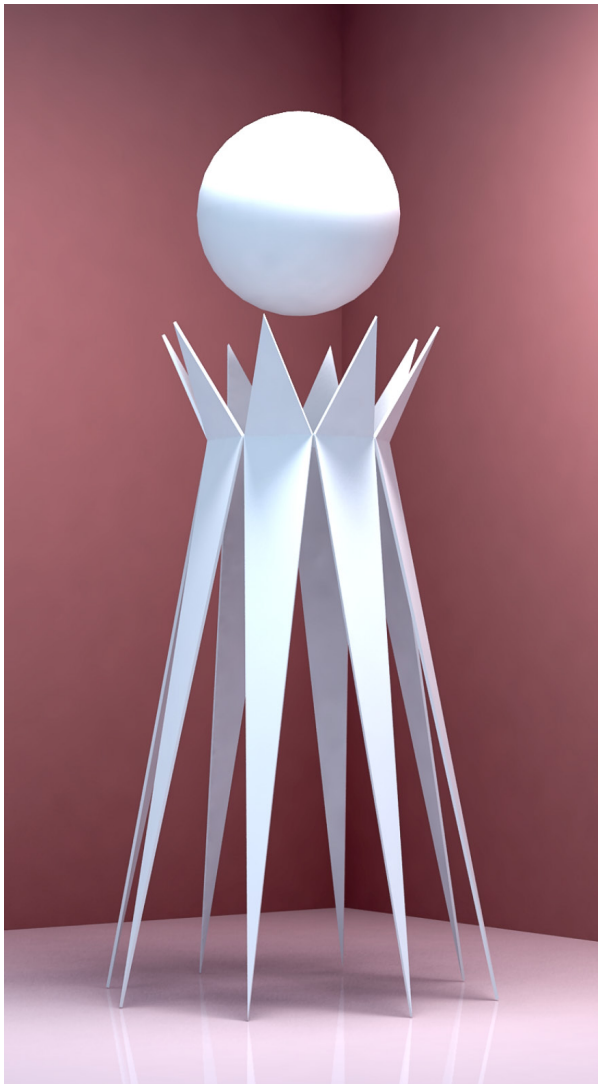 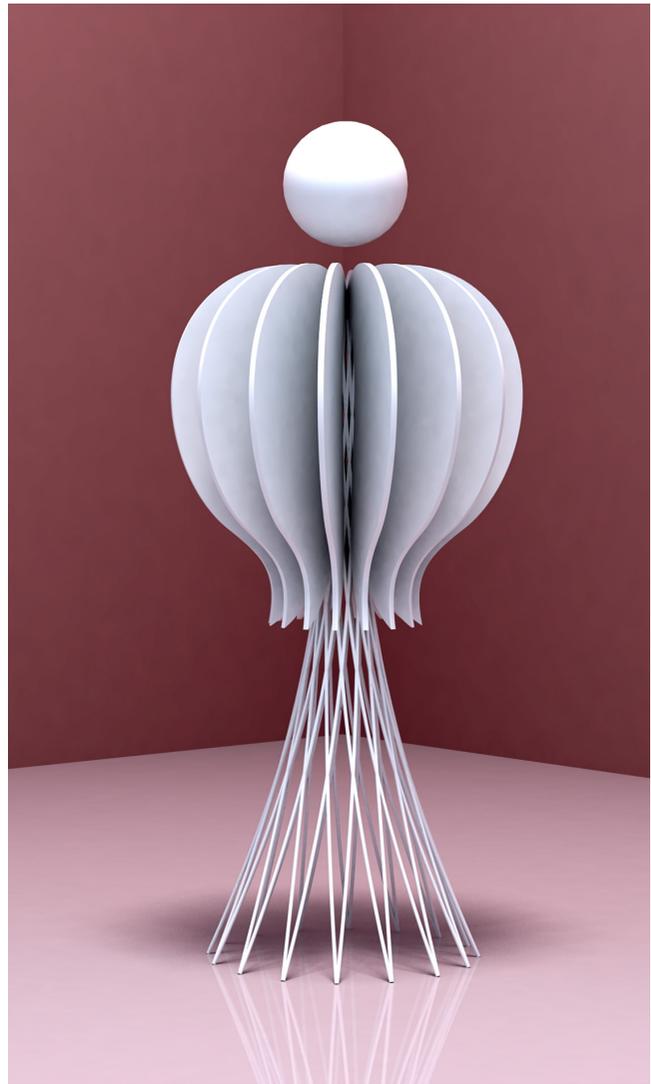

*Рис. 11. Арт-объект.* *Рис. 12. Арт-объект.*

торые в силу специфики их направленности могут помочь реанимировать идейный смысл профессии. В их ряду выделяется своим особым содержанием экподизайн. В отличие от других видов дизайна его средствами не просто осуществляется экспансия искусственных форм в социум, где потребитель будет пытаться найти с ними культурное взаимодействие. А оно возможно при условии наличия в личности соответствующей способности распредмечивать и оценивать полезные качества предметного наполнения среды.

Известно, что процесс воспитания культурного потребителя средствами дизайна - процесс не быстрый, в виду разности вкусов людей, ценностей, идеалов. С другой стороны, важен определенный фактор интереса к изделию, который не всегда проявляется в плоскости красивой и целесообразной формы. Очевидно, что существует множество других причин: культурного, национального, идеологического плана, создающих незримые барьеры для «дружбы» человека и изделия. В этом смысле особую важность приобретает возможность в рамках выразительных средств экподизайна осуществить включение в культурную среду не пассивную форму, интерес к которой, как было отмечено выше, рождается не всегда в силу объективных факторов. Необходимо создать образ активной формы, призывающий осмыслить его содержание на глубинном мировоззренческом уровне и способный эффективно содействовать воспитанию мыслящего гражданина.

Современная проектная практика в области экподизайна утвердила характерный принцип

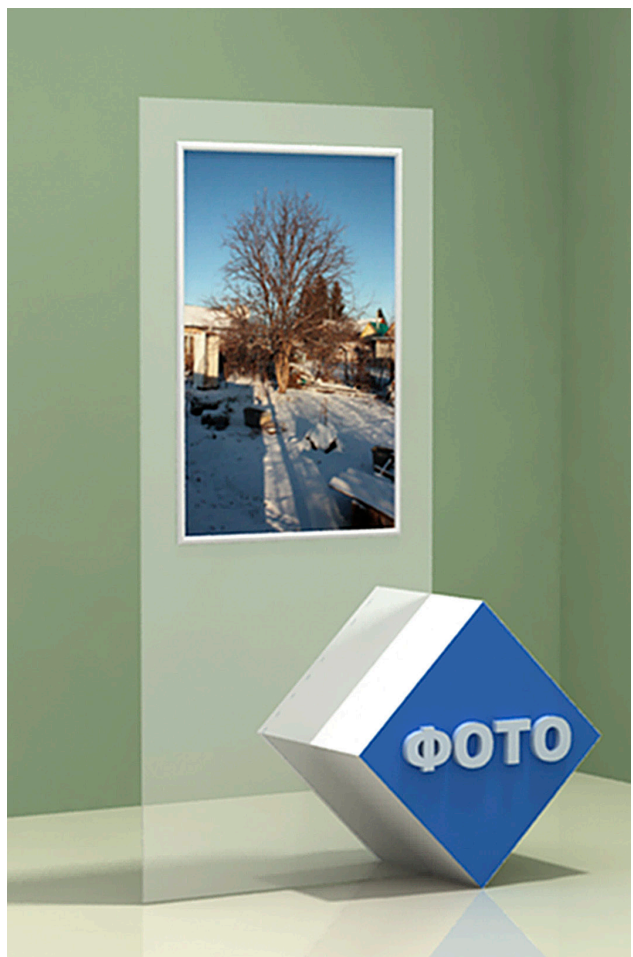 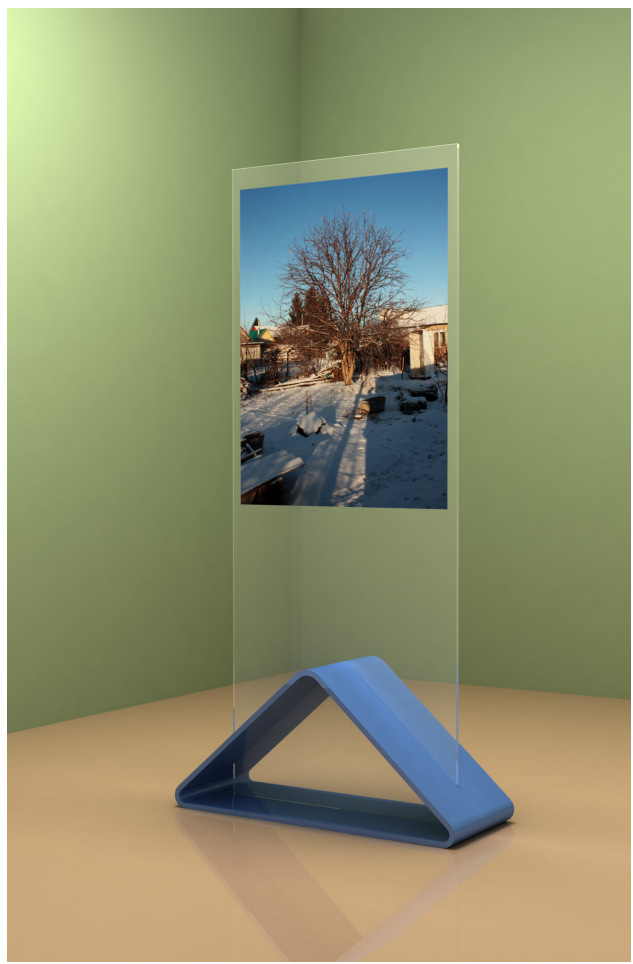

*Рис. 13, 14. Стенды для передвижных выставок. Реализована возможность оперативного монтажа и демонтажа конструктивных элементов. Изготавливаются из пластика и органического стекла.*

формирования рекламного выставочного пространства. Его смысл сводится к стандартизации конструктивных элементов и привязки мышления специалиста к определенному визуальному шаблону, насаженному модными тенденциями. Существующая подобная зависимость от навязанных стереотипов сводит форму подачи стационарных выставочных элементов к эстетической однобокости и, в результате, воспитывает шаблонный характер восприятия у зрителя. Очевидно, что этому способствует и принятая форма организации стационарных экспозиций, коммерческое содержание которых реализуется в интерьерах, сугубо адаптированных под данные мероприятия.

В эксподизайне имеется особое выставочное направление: передвижная выставка. В отличие от экспозиции стационарной выставки она не задумывается только как компонент архитектурного пространства. Её смысл заключается в том, что она может функционировать в любом предметном окружении, утверждая в каждом новом пространственно-временном контексте свои концептуальные и формально-пластические позиции. Формообразующие признаки ее компонентов не подпадают под цензуру моды, принятого стиля, архитектурно-планировочных предписаний, свойственных стационарным выставкам. В этом смысле дизайнеру предоставляется свобода в разработке эстетической концепции оптимальной формальной структуры, в которой на зрительном уровне выражалась бы актуальная социальная проблема. Подобная гибкость и характерная независимость от содержания пространства, в котором устанавливаются выставочные элементы, позволяет говорить о её самодостаточности и идейной независимости. Другим важным отличительным признаком передвижной выставки является ярко выраженная ориента-

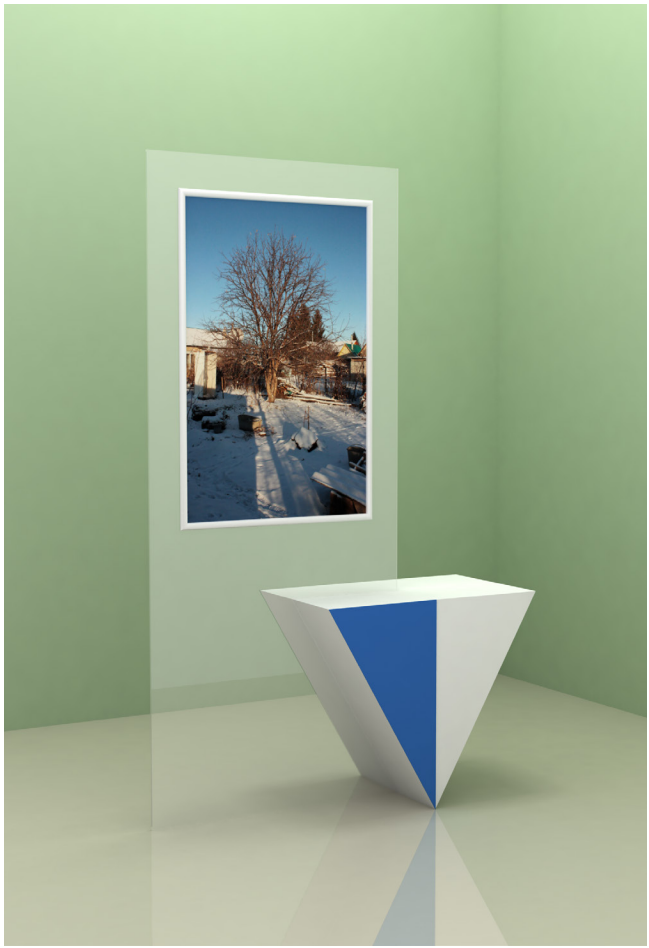 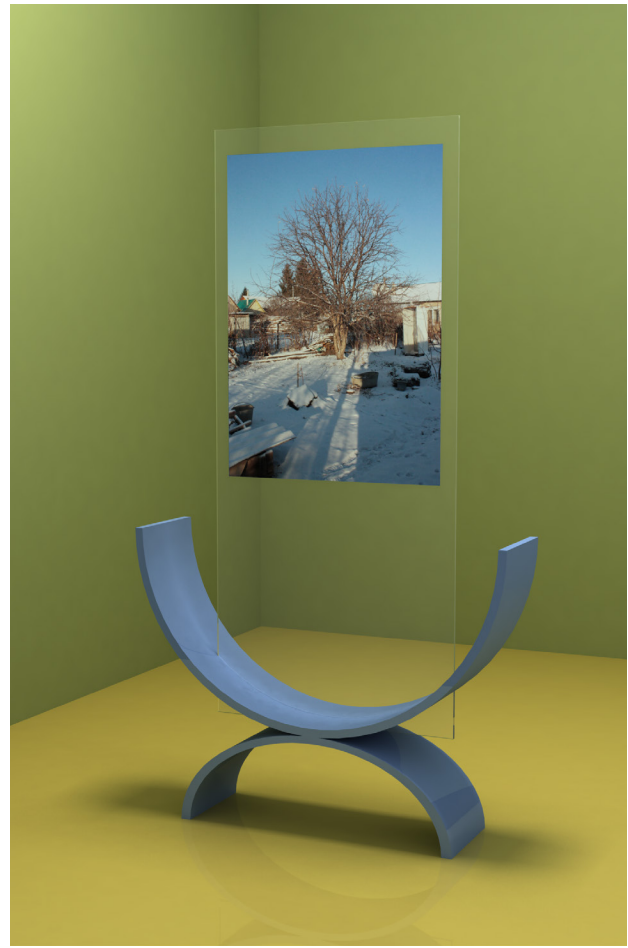

*Рис. 15, 16. Стенды для передвижных выставок. Реализована возможность оперативного монтажа и демонтажа конструктивных элементов. Изготавливаются из пластика и органического стекла.*

ция на просвещение зрителя, через включение его в проблемную ситуацию в образном её отражении. Просветительский характер такой выставки требует выработки оптимальной зрительной и технологической её модели в целях активной реализации возложенных на неё функций. Речь идёт о способности всех компонентов максимально оперативно сформировать необходимый визуальный строй и оптимально адаптироваться в любых предметно-пространственных условиях. Реализовать такое решение можно на основе оригинальной конструкторской проработки компонентов выставки, обеспечивающей оперативность сборки экспонатов, удобство транспортировки и безопасность взаимодействия с человеком. Технико-конструкторская целесообразность материальных структур позволит при необходимости осуществить количественное их воспроизводство для одновременного функционирования в разных пространственных ситуациях в целях большего охвата зрительских масс.

Известно, что в рамках выставочной экспозиции логика организации процесса личностно-предметной и межличностной коммуникации характеризует проработанную режиссуру ориентации зрителя в пространстве передвижной выставки. Учитывая, что выставочная среда передвижной выставки по отношению к внешнему окружению должна являться в определенном смысле суббпространством, необходимо ее вычленить уникальным содержанием. В противном случае среда выставки не будет восприниматься особым явлением, в котором прочитывался бы образ в знаковом его выражении. С одной стороны, уникальность характеризуется зрительной активностью, запоминаемостью формальной структуры, обеспеченные выразительностью проектного решения и образным языком подачи смысла выставочной среды средствами дизайна. С другой, показательна роль драматургии, определяющей суть ритуала, порядок соприкосновения зри-

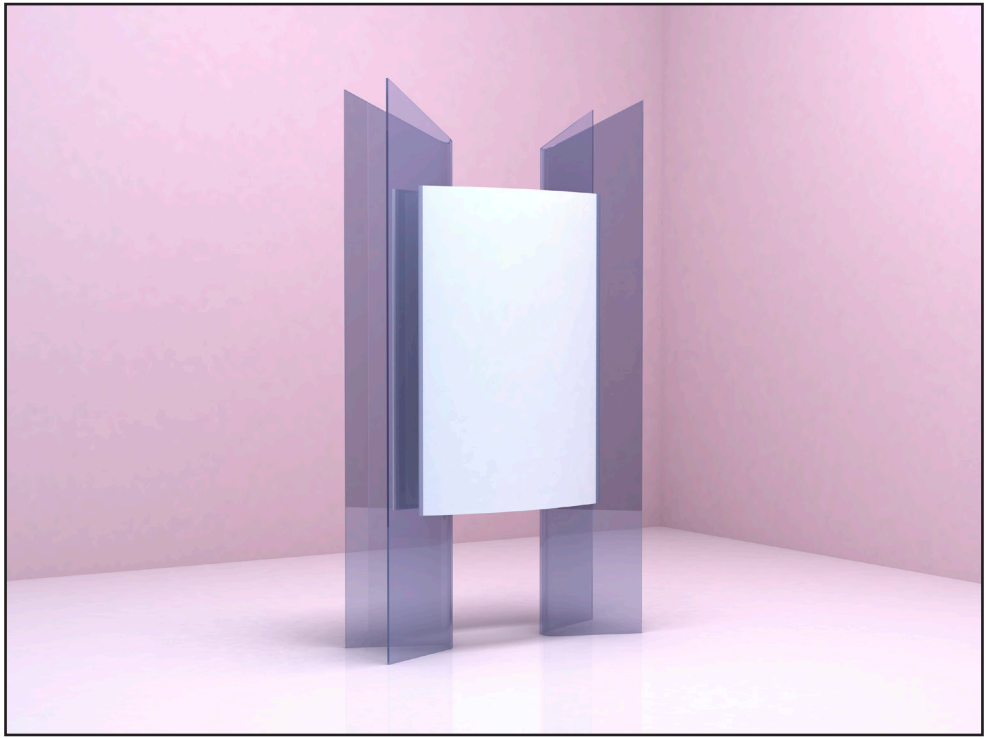

*Рис. 17. Стенд для передвижной выставки. Крепление плоскости осуществляется путем её вставки в вырез вертикальных держателей.*

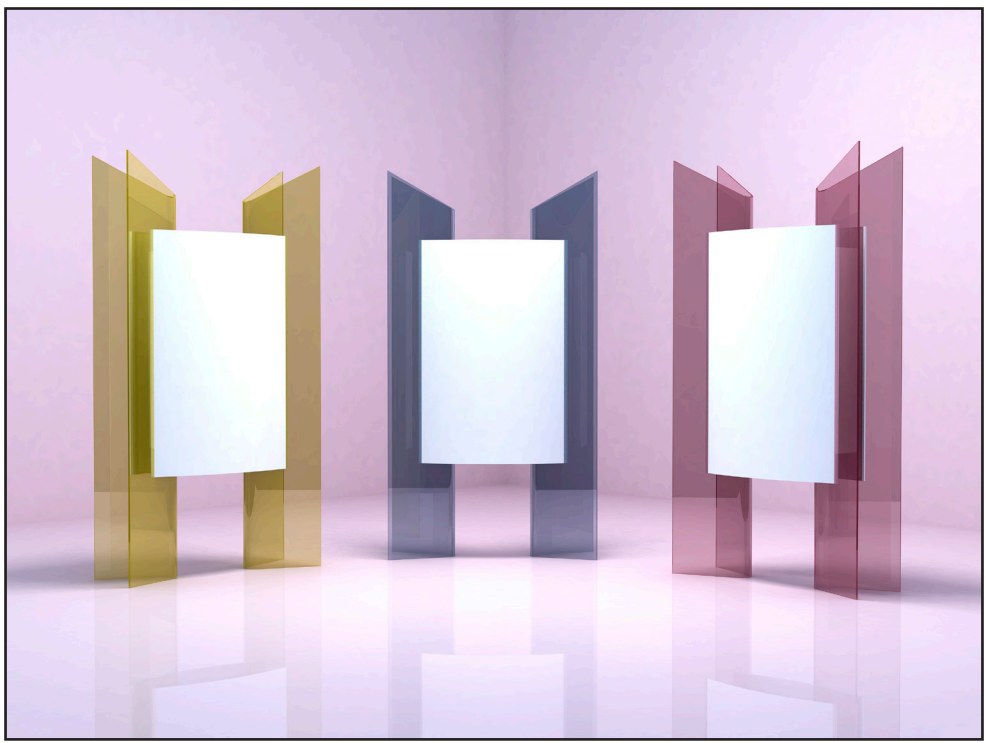

*Рис. 18. Стенды в разных цветовых исполнениях.*

теля с проблемой. Все это в совокупности должно служить цели эмоционально-эстетического воздействия на зрителя через объективные психологические закономерности пробуждения эмоциональной реакции на выраженную проблему. Очевидно, что динамика раскрытия её смысла формируется в рамках 3 уровней познания: живого созерцания, абстрактного мышления и практики. Результирующим продуктом освоения зрительного ряда и смысловой его основы станет

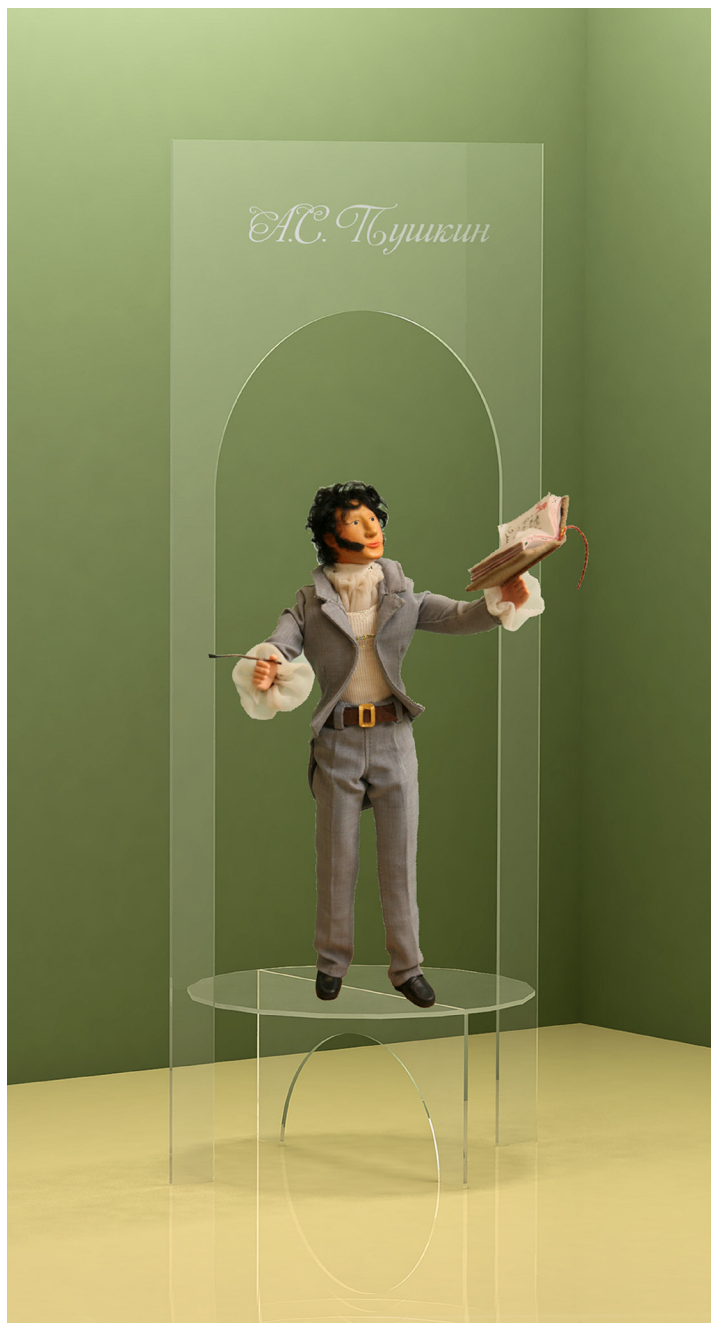

*Рис. 19.*
*Стенд для выставки авторских кукол О. Лантух.*

побуждение личности к рефлексии. В её природе выражается диалектика рационального и эмоционального аспектов познания средовой ситуации, которая должна стать определяющим принципом гармонизации отношения «зритель – выставочная среда». Соответственно, что художественно-конструкторский метод формирования проектного стиля как фактора формирования средового образа в единстве его знаковости и многозначности, а также отражение принципов технической эстетики, являются условием достижения высокого эстетического уровня выставочной среды, в обстановке которой создаётся возможность организации полноценного процесса межличностной и личностно-предметной коммуникации.

### Структурные компоненты передвижной выставки

В структурном отношении выставочная среда требует наличия таких структурообразующих элементов, которые бы формировали целостность получаемых впечатлений и не создавали зрительную неупорядоченность. В этом плане видится следующий оптимальный состав мате-

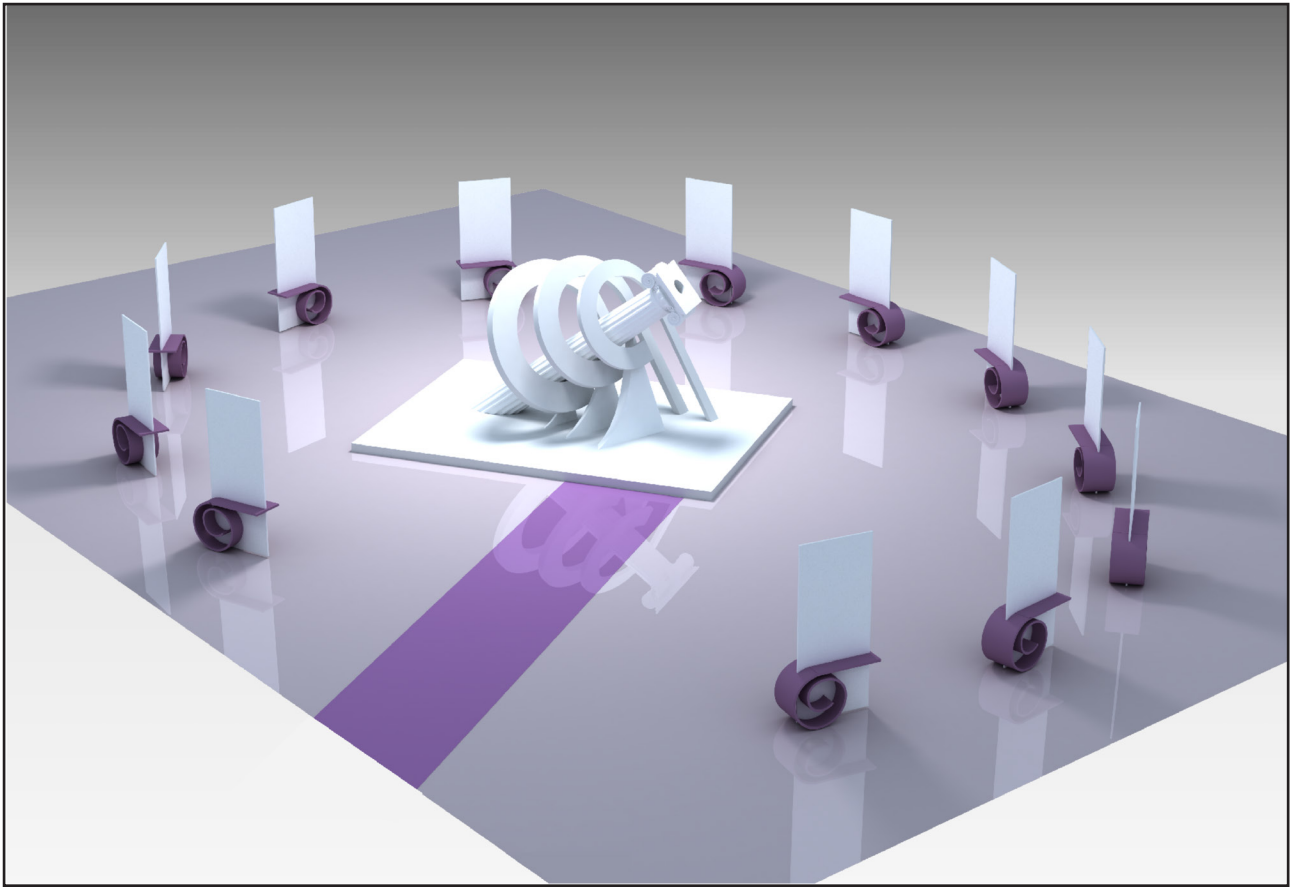

*Рис. 20. Передвижная выставка «Архитектура города». Общий вид.*

риальных компонентов выставки: арт-объект - главный композиционный центр среды и стенды для размещения информации. Оптимальность этого решения продиктована простотой их организации в пространстве и удобством соподчинения друг другу.

Арт-объект представляет собой проектное решение, сформированное средствами арт-дизайна. В этом его основное отличие от обычной инсталляции, моду на которую можно наблюдать у представителей профессионального сообщества художников, дизайнеров, скульпторов, так и обычных любителей самовыразиться, утолить назревшую потребность в творчестве. Существующая относительная свобода самовыражения в формате инсталляции делает эту потребность легко осуществимой, так как она не подпадает под цензуру строгих композиционных правил и проектных директив.

Основным признаком отличия арт-объекта от инсталляции и его видовой идентичностью является ярко выраженный дизайнерский способ формообразования, основанный на грамотном структурировании элементов в рамках требований объемно пространственной композиции. Также в арт-объекте, как и во всех продуктах, сформированных художественно-конструкторским методом, показательна неизобразительность интерпретации проблемы в отличие от свободной трактовки формы, принятой в инсталляции. Средствами арт-объекта выражается эстетический образ проблемы, создаётся зрительный акцент в пространстве, в рамках которого зритель сможет завершить процесс культурного диалога и, в идеале, выполнить философское обобщение. В этом отношении, непрямолинейный язык изложения проблемы в арт-объекте является условием её домысливания уже на уровне многозначности образа выставки вследствие целостного воздействия средовой ситуации под эгидой всех ее атрибутов.

Таким образом, арт-объект как центральный в структурном отношении элемент передвижной выставки определяет процесс раскрытия подтекста выражаемой проблемы через меха-

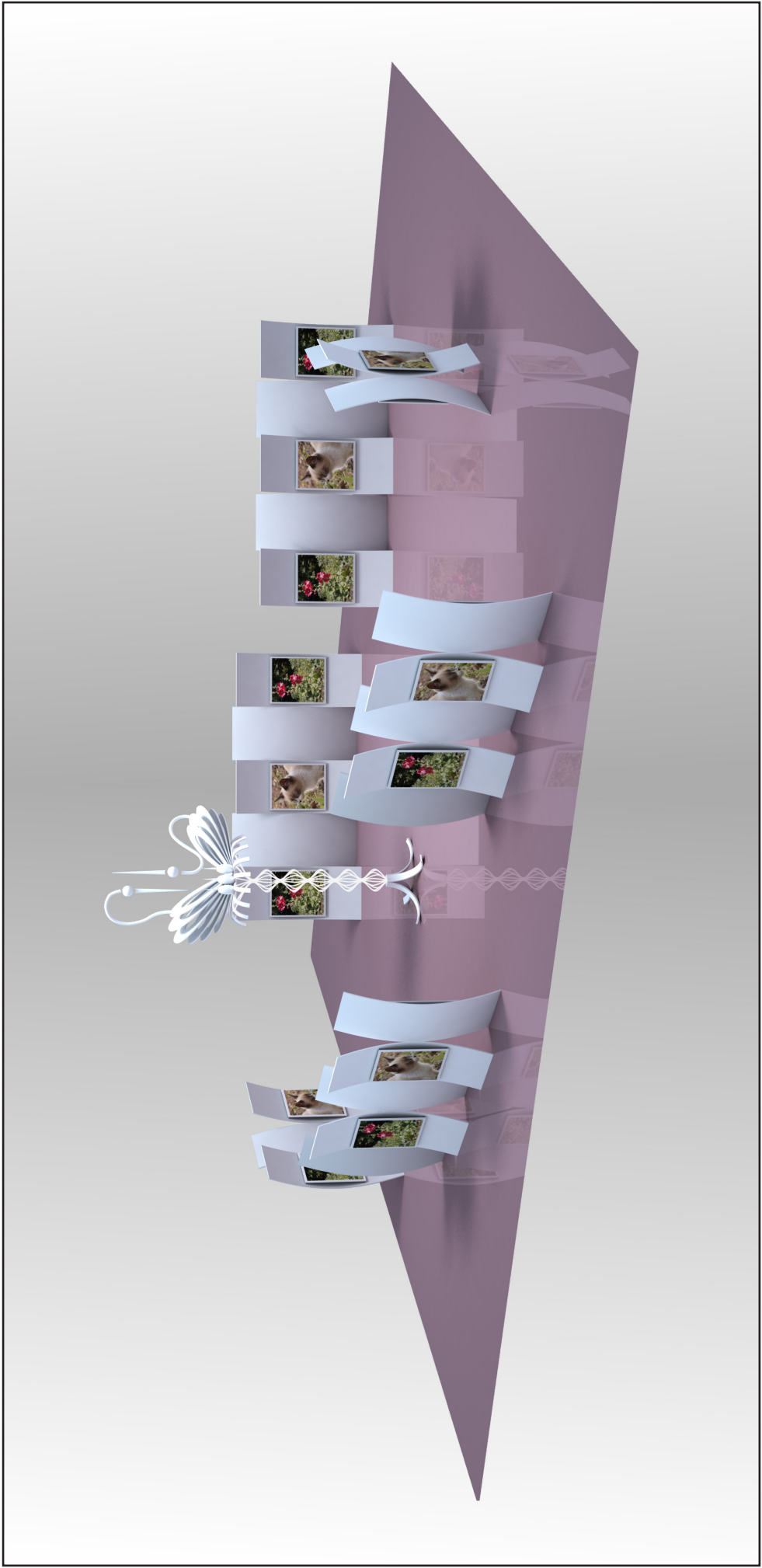

Рис. 21. Передвижная выставка «Мир живой природы».
Общий вид.

низм проектной модели структурного воздействия на сознание зрителя.

На рис. 4 показан арт-объект, зримо олицетворяющий тему передвижной выставки «Мир живой природы». Представлена интерпретация красоты фауны в образе журавлиной семьи. В рамках поставленной задачи было найдено проектное решение. Оно материализовалось в объемно-пространственной структуре, атрибуты которой формируют ясно читаемый сюжет, воплощенный дизайнерским способом формообразования. Важно отметить, что в идеале удобочитаемость визуального ряда должна распространяться на все категории зрителей, с учетом их социальной неоднородности и возрастных различий. Для дизайнера подобная задача обеспечения свободной расшифровки образных характеристик арт-объекта зрителем на уровне сенсорного приема информации, связана с постоянным совершенствованием профессиональных навыков. Выразительные средства дизайна позволяют добиться композиционной цельности, стилистического единства, пластической выразительности всего арт-объекта. Его композиции, составленной из абстрактных геометрических элементов, свойственен неизобразительный визуальный язык повествования. Такая форма изложения является критерием дизайнерского метода преобразования вещества природы. Объективно, что содержание категориального аппарата художественного конструирования не допускает возникновения пластического схода в сторону украшательства. В этой связи дизайнеру важно определить образные рамки пересказа и тем самым не нарушить меру условности, показательную для состоятельности эстетического образа.

В рамках поставленной задачи было решено отказаться от использования метода «цветовой коррекции и преобразования формы» арт-объекта в виду достигнутой пластической самодостаточности. В других случаях использование цвета может быть оправданным решением в плане выделения смысловой доминанты в композиционных элементах арт-объекта, исправления дисгармоничных отношений объемов, усиления эмоционально-образного впечатления. Вместе с тем известно, что гармоничная форма не нуждается во внешних дополнениях. В искусстве есть множество свидетельств преимущества завершенной формы. Ярким примером единства содержания и формы, без нанесения ненужной бутафории, может служить статуя Поликлета «Дорифор». Поэтому не случайно, что полноценная геометрия объема объекта в единстве его эстетических и технических характеристик является мерилом высокой пластической культуры и чистоты формы еще со времен Античности. Это положение актуально и для дизайна. В этом отношении, принцип «самодостаточной формы» должен стать ведущей составной частью проектной идеологии дизайнера в целях утверждения объективного смысла визуализированной проблемы в сознании личности.

Любое мероприятие, организованное средствами дизайна, и обращенное к зрительской аудитории, является своеобразной проверкой специалиста на широту его компетентности. Однако проектная деятельность может вызвать у дизайнера искушение соответствовать только нормам формального мастерства и абсолютизировать техническую сторону профессии, что может сузить сферу восприятия содержания общественных проблем до рамок, ограниченных ремесленнической однобокостью. Поэтому, функционирование дизайнера в пространстве проектной культуры подразумевает его проявление как специалиста, мыслящего и категориями духовно-нравственной практики для гармонизации всех сторон жизнедеятельности человека. Данный подход подразумевает глубокое изучение имеющейся проблемы в плане ее философской, нравственной и социальной противоречивости и материализацию авторского отношения к ней соответствующими средствами. В этом смысле культура проектирования среды передвижной выставки должна отвечать вышеназванным требованиям. В противном случае будет недостижим воспитательный и духовно-нравственный эффект от восприятия выставочных компонентов, потенциальные возможности которых ограничены методом «эстетизации внешнего слоя формы». В результате формируется примитивность её внутренней сути.

Рисунок 5 демонстрирует пример арт-объекта, эстетика которого утверждает целесообразность и социальную значимость материальной структуры с ярко выраженной проблематизацией формы. Тема выставки связана с сохранением классического архитектурного наследия в проти-вовес современному новострою с примитивным формообразованием. В качестве символа, олицетворяющего прогрессивный идеал архитектурной формы, применена колонна с ионическим ордером. Геометрические свойства ордера перенесены и на элементы композиции, выражающие атаку бессодержательной архитектуры на традиционные архитектурные ценности. Соблюдение данных закономерностей является условием формирования стилистически однородного пластического образа. Объемно пространственная структура арт-объекта характеризуется ярко выраженной ассиметричной компоновкой. Следует отметить, что задача формирования главного смыслового компонента передвижной выставки как средства, интегрирующего в своем образе силу идеи и адекватную ей объемную пластику, предъявляет к дизайнеру более жесткие требования в плане знаний закономерностей формообразования. Все это свидетельствует о том, что передвижная выставка в своем системном содержании является площадкой совершенствования проектных навыков дизайнера, а также его интеллектуальной, нравственной, духовно-мировоззренческой формовкой. Очевиден в этом процессе диалогический способ преобразования личности дизайнера и зрителя. Он сводится к взаимообусловленности качественных изменений в мышлении: когда трансформация эмоциональных впечатлений в соответствующие умозаключения и дальнейшее обретение новых жизненных стимулов у зрителя, создает основу для усиления профессиональной мотивации дизайнера к решению задач на более высокой ступени профессионального мастерства.

На рисунах 9. 10, 11, 12 представлены образцы арт-объектов, спроектированные для функционирования в среде передвижной выставки соответствующей тематической направленности. В качестве основного материала для их изготовления выбран пластик, который позволяет добиться высокой культуры сборки и чистоты визуально-эстетического эффекта. Конструкция арт-объектов и их объемно-пластические характеристики воплощают художественно-конструкторский метод создания лаконичного и выразительного эстетического образа.

Известно, что тематизация выставочного пространства осуществляется совокупностью информационных носителей, в ряду которых выставочный стенд является держателем и демонстратором наглядностей. Его геометрия, формообразующие и размерные свойства должны быть подчинены функции «зонирования пространства» согласно требованиям эргономики для обеспечения положительного психо-эмоционального состояния человека. С другой стороны, эстетика выставочного стенда является значимым фактором создания уникальности зрительного образа выставочного пространства и зримым признаком организованности в нём структурных отношений. Речь идёт о том, что выставочные стенды структурируют среду в вертикальной плоскости. Тем самым они выявляют важное структурное отношение вертикальной пространственной оси к поверхности пола.

Выставочный стенд относится к средовым объектам. В рамках функционирования в среде передвижной выставки в нем должны быть воплощены технико-конструктивные свойства, обеспечивающие легкость, безопасность, оперативность монтажа и демонтажа, удобство транспортировки. Технические характеристики стенда должны решать задачу демонстрации и организации наглядного материала с учетом его свойств. В этом смысле логика формообразования компонентов выставочного стенда складывается в соотнесенности с признаками выставочных экспонатов.

Выставочные стенды, реализующие в своем содержании общественно-преобразующие и воспитательные цели передвижной выставки, необходимо наделять такими качествами как академичность и представительность. Именно эти показатели отличают носителей экспонатов передвижной выставки от стендов коммерческой направленности. Однако при необходимости, оригинальные эстетические и конструктивные свойства формы стендов могут быть ориентированы и на коммерческие интересы, так как высокая культура формообразования продукта дизайна способна выполнять многие общественно-полезные функции и отвечать разным интересам.

Рисунки с 13 по 18 демонстрируют примеры выставочных стендов, простых по конструкции и, вместе с тем, выразительных по форме. Их целесообразность раскрывается в способности формировать нужную пространственную ситуацию. Например, конструктивные характеристики стендов, показанных на рисунках 13, 14, 15, 16, позволяют сочленять их друг с другом с помощью обычной вставки плоскости в паз её держателя и создавать непрерывный ряд наглядного материала. Такое же удобство монтажа реализовано в конструкции стендов, показанных на рисунках 17, 18. На рисунке 19 показан стенд для демонстрации объемных экспонатов. В данном случае на стенде размещаются авторские куклы. Конструкция стенда цельная, плоскости-держатели экспонатов складывающиеся. Стенд выполнен из оргстекла, прозрачность которого тонко дополняет и содействует раскрытию кукольного образа.

Все стенды, приведенные в качестве примеров, изготавливаются из пластика и органического стекла, и позволяют демонстрировать плоский наглядный материал с двух сторон. Также элементы стендов могут по-разному обыгрываться цветом и графикой. Дизайнеру в этом смысле важно все возможные выразительные средства оправданно применить для организации плодотворного процесса соприкосновения зрителя с проблемой.

На рисунках 20 и 21 показаны проектные предложения передвижных выставок. Очевидно, что утверждаемая их средствами проблема определяет формирование уникального зрительного образа, который в свою очередь является критерием соответствия формообразующих, объемно-пластических характеристик компонентов содержанию темы выставки.

*Заключение*

Представленные примеры проектных предложений передвижных выставок являются одними из возможных решений задачи целенаправленного преобразования внутреннего мира человека средствами эксподизайна. Арсенал его выразительных средств содержит неисчерпаемые воспитательные, психологические, эстетические и духовные возможности приобщения личности к процессу диалоговой формы восприятия, осмысления и оценки имеющейся проблемы. Подобный стиль культурного взаимодействия, осуществляемого в организованном и оптимизированном выставочном пространстве, утверждает общественную значимость передвижной выставки как одной из возможных, но эффективных форм реализации принципов гражданского университета.

В связи с вышесказанным подчеркнем, что дизайнеру в этом диалоге определена значимая роль. Она требует от него высокой степени моральной ответственности, издревле присущей деятелям искусства с гуманистическим стилем мышления, а также миссионерским восприятием мира в лучших традициях классического отечественного меценатства.